\documentclass[preprints,review,accept,pdftex,oneauthors]{mdpi} 

\firstpage{1} 
\pubvolume{1}
\issuenum{1}
\articlenumber{0}
\pubyear{2023}
\copyrightyear{2023}
\datereceived{ } 
\daterevised{ } % Comment out if no revised date
\dateaccepted{ } 
\datepublished{ } 
\hreflink{https://doi.org/} % If needed use \linebreak
\Title{X-Ray Tests of General Relativity with Black Holes}

\TitleCitation{X-Ray Tests of GR with Black Holes}

\Author{Cosimo Bambi \orcidA{}}

\AuthorNames{Cosimo Bambi}

\AuthorCitation{Bambi, C.}
\address{%
Center for Field Theory and Particle Physics and Department of Physics, Fudan University\\
200438 Shanghai, China; bambi@fudan.edu.cn}

\abstract{General relativity is one of the pillars of modern physics. For decades, the theory has been mainly tested in the weak field regime with experiments in the Solar System and radio observations of binary pulsars. Until 2015, the strong field regime was almost completely unexplored. Thanks to new observational facilities, the situation has dramatically changed in the last few years. Today we have gravitational wave data of the coalesce of stellar-mass compact objects from the LIGO-Virgo-KAGRA Collaboration, images at mm wavelengths of the supermassive black holes in M87$^*$ and Sgr~A$^*$ from the Event Horizon Telescope Collaboration, and X-ray data of accreting compact objects from  a number of X-ray missions. Gravitational wave tests and black hole imaging tests are certainly more popular and are discussed in other articles of this Special Issue. The aim of the present manuscript is to provide a pedagogical review on X-ray tests of general relativity with black holes and to compare this kind of tests with those possible with gravitational wave data and black hole imaging.}


\begin{document}

%%%%%%%%%%%%%%%%%%%%%%%%%%%%%%%%%%%%%%%%%%

\section{Introduction}

The theory of general relativity is one of the pillars of modern physics. The theory was proposed by Einstein at the end of 1915~\cite{Einstein:1916vd} and in more than 100~years it has passed a large number of observational tests without requiring any modification from its original version~\cite{Will:2014kxa}. The first test of general relativity can be dated back to 1919, when Eddington and collaborators measured the effect of light bending by the Sun during a Solar eclipse~\cite{Dyson:1920cwa}. After that observation, Einstein and his theory became soon very popular, but the precision and the accuracy of that measurement were actually quite poor and that experiment was not really able to distinguish general relativity from alternative scenarios. Systematic tests of general relativity started only in the 1960s with experiments in the Solar System and in the 1970s with accurate radio observations of binary pulsars~\cite{Will:2014kxa}. Solar System experiments and radio pulsar observations can mainly test the {\it weak field regime}, where the astrophysical system can be described as a Newtonian system plus some small corrections. With those tests, we want to measure such small corrections and check whether they are consistent with the predictions of general relativity. In the past 20~years, there have been significant efforts even to test general relativity on {\it large scales} (galactic scales or above) in response to the problems of dark matter and dark energy~\cite{Jain:2010ka,Koyama:2015vza,Ferreira:2019xrr}. Up to some years ago, the {\it strong field regime} was almost completely unexplored. However, since 2015 the situation has dramatically changed. Today we can probe the strong field regime with gravitational wave data from the LIGO-Virgo-KAGRA Collaboration (see, e.g., Refs.~\cite{LIGOScientific:2016lio,Yunes:2016jcc,LIGOScientific:2019fpa}), images of the supermassive black holes in M87$^*$ and Sgr~A$^*$ from the Event Horizon Telescope Collaboration (see, e.g., Refs.~\cite{Bambi:2019tjh,EventHorizonTelescope:2020qrl,Vagnozzi:2022moj}), and X-ray data from a number of X-ray missions~(see, e.g., Refs.~\cite{Cao:2017kdq,Tripathi:2018lhx,Tripathi:2020yts}).

Black holes are ideal laboratories to test the strong field regime as they are the sources of the strongest gravitational fields that can be found today in the Universe~\cite{Bambi:2015kza,Bambi:2017khi}. In 4-dimensional general relativity, black holes are relatively simple objects and are completely described by a few number of parameters. This is the celebrated result of the {\it no-hair theorem}, which is actually a family of theorems with different versions and a number of extensions~\cite{Chrusciel:2012jk}. According to the no-hair theorem, astrophysical black holes should be characterized only by their mass, spin angular momentum, and electric charge. Since the electric charge is normally negligible for macroscopic astronomical objects, the mass and the spin should be the only two parameters relevant for astrophysical black holes and the spacetime metric should be described by the Kerr solution~\cite{Kerr:1963ud}\footnote{Deviations from the Kerr metric due to the presence of an accretion disk, nearby stars, or a non-vanishing equilibrium electric charge can be estimated, but normally they turn out to be completely negligible; see, e.g., Refs.~\cite{Bambi:2017khi,Bambi:2017iyh,Bambi:2014koa}. For example, a non-vanishing equilibrium electric charge is a natural consequence of the difference between the values of the masses of electrons and protons as well as of their different photon scattering cross-sections. However, the value of the equilibrium electric charge turns out to be completely negligible for the spacetime geometry around a black hole; for more details, see, for instance, Section~6.5.2 in Ref.~\cite{Bambi:2017khi}.}. However, such a conclusion can be easily invalidated in extensions of general relativity (see, e.g.,~\cite{Kleihaus:2011tg}), if macroscopic quantum gravity effects show up in the vicinity of these compact objects (see, e.g.,~\cite{Giddings:2017jts,Mottola:2023jxl}), or in the presence of exotic matter fields (see, e.g.,~\cite{Herdeiro:2014goa,Herdeiro:2016tmi}).

There are already some recent reviews on tests of general relativity with black hole X-ray data; see, for instance, Ref.~\cite{Bambi:2022dtw}, where the interested reader can find more details and a long discussion on the systematic uncertainties of this kind of measurements. The aim of the present manuscript is to provide a shorter and more easily-accessible review on the topic. In particular, Section~\ref{sec-sbhs} shows the physics involved in the calculations of synthetic black hole spectra and how we can modify these calculations to test new physics.
While X-ray tests are not as popular as tests with gravitational wave data and black hole imaging, they can provide very competitive constraints, at least comparable and complementary to those from gravitational waves and certainly stronger than those from black hole imaging.

%%%%%%%%%%%%%%%%%%%%%%%%%%%%%%%%%%%%%%%%%%

\section{Disk-Corona Model}

To test general relativity with black hole X-ray data, we need to study very special systems. To have a rough idea of how unique these systems are, we can consider the stellar-mass black holes in our Galaxy. From stellar population and evolution studies, we expect that there are $10^8$-$10^9$ stellar-mass black holes formed from the collapse of heavy stars in our Galaxy~\cite{Timmes:1995kp}. On the other hand, we know less than 100~stellar-mass black holes that (more often sporadically for short times, from weeks to months) have an accretion disk sufficiently bright to be studied with our X-ray observatories. Some of these black holes have been studied well and we have hundreds of observations from different X-ray missions, while for other objects we have only a few observations. Among thousands of X-ray observations of stellar-mass black holes in our Galaxy, we have to select the sources and the observations suitable for testing general relativity, and in the end we find that there are only a few spectra from the most recent X-ray missions that can do the job!

\begin{figure}[H]
\centering
\includegraphics[width=10.5 cm]{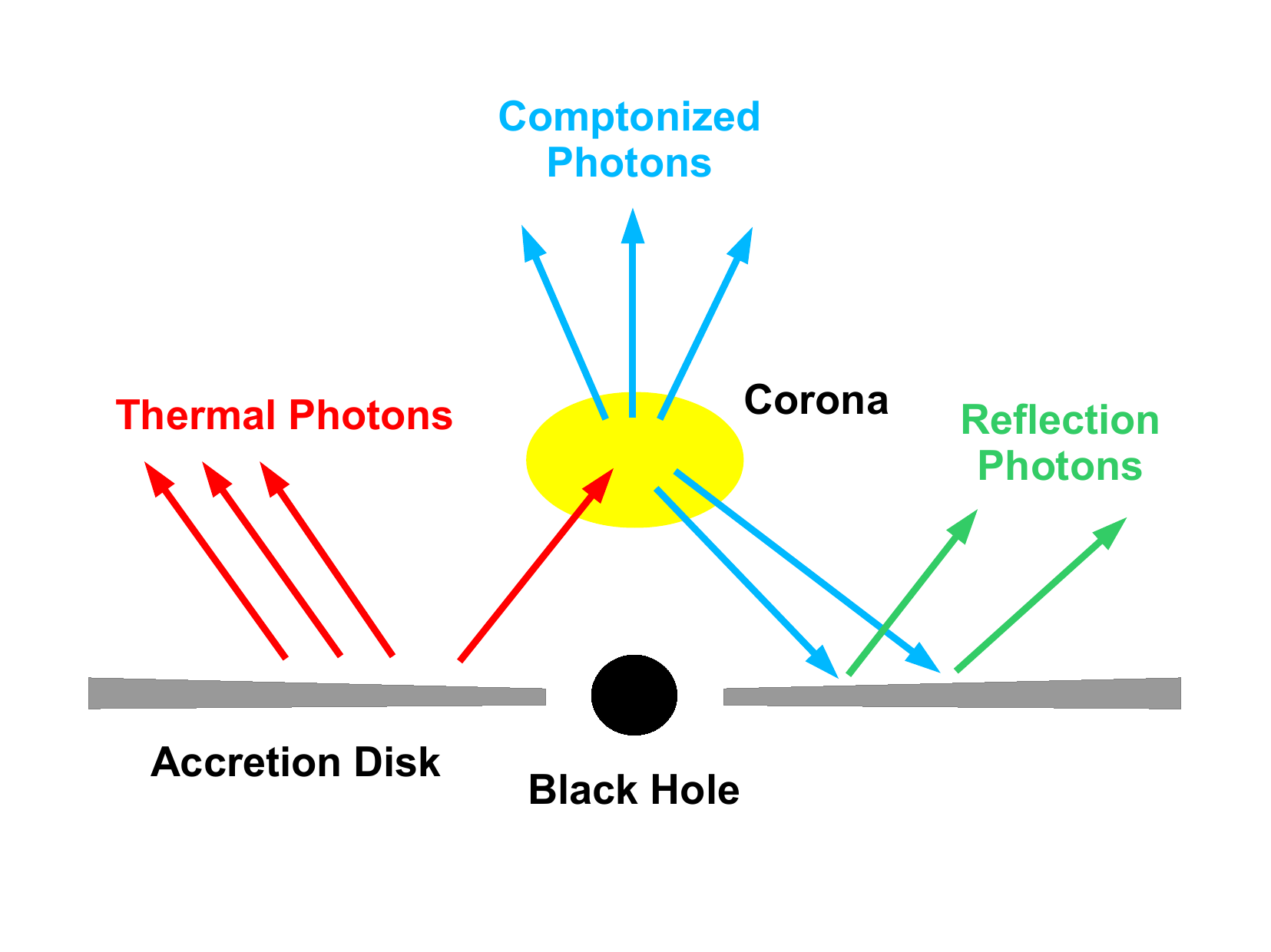}
\vspace{-0.8cm}
\caption{Disk-corona model. The black hole is surrounded by a {\it cold}, geometrically thin, and optically thick accretion disk. The corona is some {\it hot} gas near the black hole and the inner part of the accretion disk. Thermal photons from the disk inverse Compton scatter off free electrons in the corona. Some Comptonized photons illuminate the accretion disk and produce the reflection spectrum. Figure from Ref.~\cite{Bambi:2021chr} under the terms of the Creative Commons Attribution 4.0 International License. \label{f-corona}}
\end{figure}

The prototype of astrophysical system for our tests is shown in Fig.~\ref{f-corona} (for more details, see~\cite{Bambi:2020jpe} and references therein). The black hole can be either a stellar-mass black hole in an X-ray binary system or a supermassive black hole in an active galactic nucleus. This black hole is surrounded by a {\it cold}, geometrically thin, and optically thick accretion disk. Such a disk forms when the angular momentum of the accreting material is high and the mass accretion rate is between a few percent and about 30\% of the Eddington limit of the central object. In these conditions, every point on the surface of the accretion disk emits a blackbody-like spectrum and the whole disk has a multi-temperature blackbody-like spectrum, because the temperature of the accreting material increases as the radial distance from the black hole decreases. The thermal spectrum of the disk is normally peaked in the soft X-ray band (0.1-10~keV) for stellar-mass black holes in X-ray binary systems and in the UV band (1-100~eV) for supermassive black holes in active galactic nuclei. The ``corona'' is some {\it hot} ($\sim 100$~keV) gas around the black hole and the inner part of the accretion disk: the corona may be the base of the jet, some atmosphere above the accretion disk, hot material in the plunging region between the inner edge of the accretion disk and the black hole, etc. Since the disk is cold and the corona is hot, thermal photons from the accretion disk can inverse Compton scatter off free electrons in the corona. These Comptonized photons have a spectrum that can be normally approximated well by a power law with an exponential high-energy cutoff. Such a power law component is often the dominant component in the X-ray spectra of accreting black holes. Some Comptonized photons can illuminate the accretion disk: Compton scattering and absorption followed by fluorescent emission generate the so-called reflection spectrum.

The reflection spectrum in the rest-frame of the material of the disk is characterized by narrow fluorescent emission lines in the soft X-ray band and a Compton hump with a peak at 20-30~keV~\cite{Ross:2005dm,Garcia:2010iz}. The most prominent emission line is normally the iron K$\alpha$ line, which is at 6.4~keV in the case of neutral or weakly ionized iron atoms and shifts up to 6.97~keV in the case of hydrogen-like iron ions. The reflection spectrum of the disk observed far from the source is blurred by relativistic effects (gravitational redshift and Doppler boosting)~\cite{Laor:1991nc,Bambi:2017khi}.

%%%%%%%%%%%%%%%%%%%%%%%%%%%%%%%%%%%%%%%%%%

\section{Synthetic Black Hole Spectra}\label{sec-sbhs}

\begin{figure}[H]
\centering
\includegraphics[width=10.5 cm]{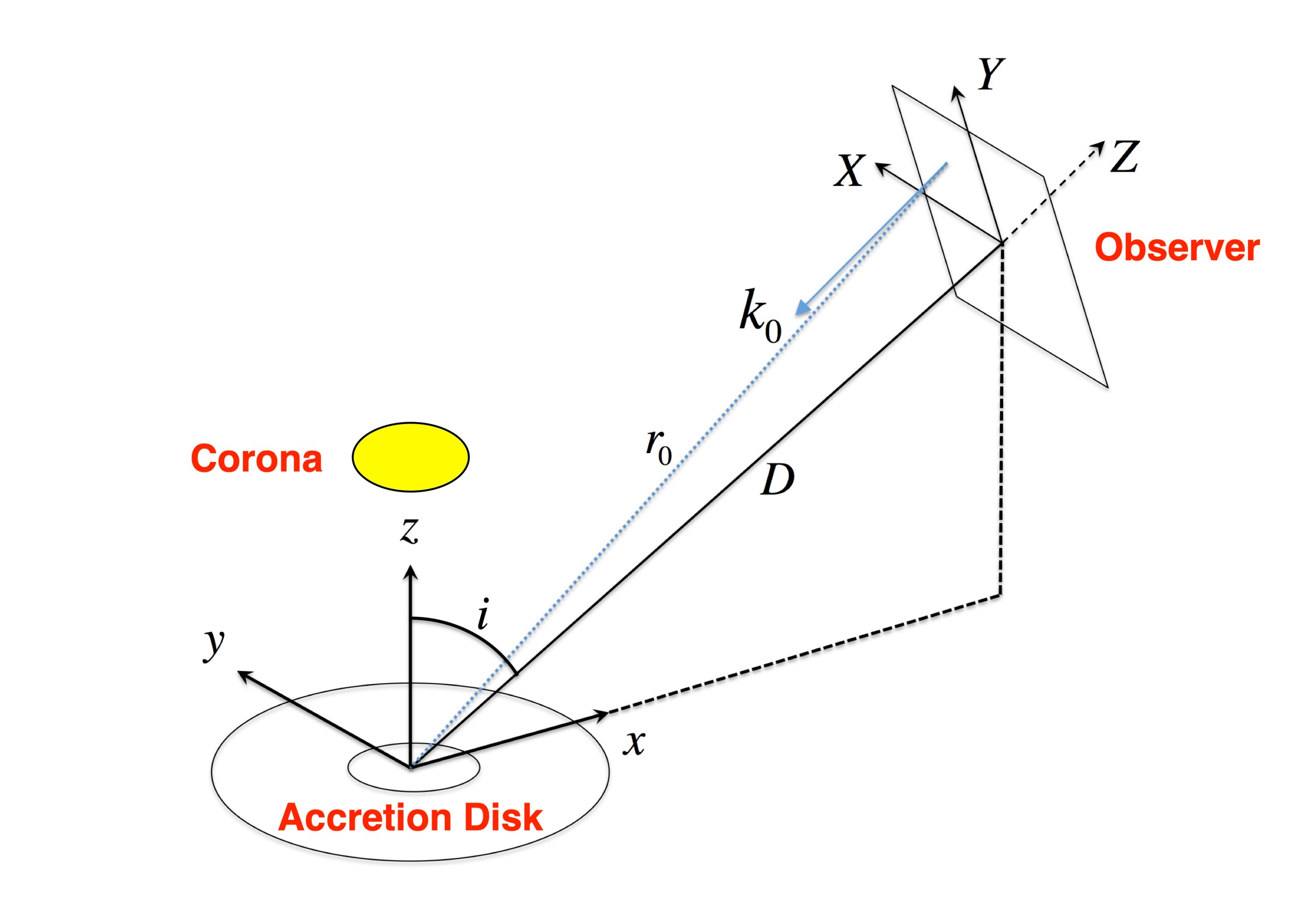}
\vspace{-0.1cm}
\caption{Set-up to calculate the spectrum of the accretion disk observed far from the source. See the text for more details. Figure adapted from Ref.~\cite{Bambi:2016sac}. \label{f-setup}}
\end{figure}

In order to understand how we can use black hole X-ray data to test fundamental physics, it can be useful to briefly review how we calculate synthetic spectra of accreting black holes. We can consider the set-up illustrated in Figure~\ref{f-setup}, where we have the distant observer, the accretion disk, and the corona. These calculations are extensively discussed in the literature; see, for example, \cite{Bambi:2017khi} and references therein. In this section, I only outline the basic steps, without showing any formula.

The first step is to calculate the {\it redshift image} of the accretion disk. We fire photons from the image plane of the distant observer backward in time to the accretion disk. When a photon hits the disk, we calculate its redshift $g = E_{\rm o}/E_{\rm e}$, where $E_{\rm o}$ is the photon energy measured at the detection point in the rest-frame of the distant observer and $E_{\rm e}$ is the photon energy at the emission point in the rest-frame of the gas in the disk. To do this, we have to:
\begin{enumerate}
\item solve the equations of motion of the photons (the geodesic equations if we are considering a metric theory of gravity~\cite{Will:2014kxa}) in order to connect every point of the observer's image of the disk to its emission point on the disk and calculate the photon 4-momentum at the emission point;
\item calculate the motion of the material in the disk (which should depend on the accretion disk model and on the equations of motion of the particles of the disk) in order to determine the gas 4-velocity at the emission point (which, together the photon 4-momentum at the emission point determined at point~1 above, is used to calculate the photon energy at the emission point in the rest-frame of the gas in the disk, $E_{\rm e}$).
\end{enumerate}
The details of these calculations can be found, for example, in Refs.~\cite{Bambi:2017khi,Bambi:2016sac}. At the end of the calculations, one has the redshift image of the accretion disk. If the photon's equations of motion are independent of the photon energy (as it is in general relativity and any other metric theory of gravity), every point of the observer's image of the accretion disk has a well defined redshift (if not, we will have that every point of the observer's image of the accretion disk has a redshift that depends on the photon energy).

The second step is to calculate the emissivity profile of the disk. In the case of the thermal component, current models employ Novikov-Thorne disks~\cite{NT73,Page:1974he} and the emissivity profile is determined by the spacetime metric and the mass accretion rate. In the case of the reflection component, the emissivity profile is determined by the geometry and the emissivity of the corona. If we know these properties of the corona, we can fire photons from the corona to the accretion disk and calculate the emissivity profile of the reflection spectrum. These calculations are similar to those at step~1 and require to solve the equations of motion of the photons from the corona to the disk and to determine the motion of the material in the disk. Unfortunately, the properties of the corona are not yet well understood. Current reflection models often do not assume a specific coronal geometry and employ some phenomenological emissivity profiles (like a power law, broken power law, or twice broken power law) that are thought to be able to approximate well the emissivity profile generated by any possible coronal geometry.

The third and last step is to calculate the spectrum at the emission point in the rest-frame of the material of the disk. In the case of the thermal component, the spectrum is simply a blackbody spectrum with possible corrections due to the disk atmosphere (mainly photon-electron scattering). In the Novikov-Thorne model, the temperature at every radius of the disk is determined by an equation that follows from the conservation of mass, energy, and angular momentum and depends on the spacetime metric and the mass accretion rate. In the case of the reflection spectrum, we have to solve radiative transfer equations and the calculations involve atomic physics and some assumptions about the structure of the accretion disk. In general relativity and in any other metric theory of gravity, the atomic physics in the strong gravitational field around a black hole is the same as the atomic physics in our laboratories on Earth. This is because locally the laws of non-gravitational physics are those of special relativity in all metric theories of gravity. However, this is not the case in some theories of gravity where the values of some fundamental constants (e.g., the fine structure constant $\alpha$, the electron mass $m_e$, etc.) can be different in strong and weak gravitational fields; see, e.g., Refs.~\cite{Davis:2016avf,Bambi:2022dtw,Bambi:2022lhq}. This may have an impact on the atomic energy levels, photon-electron scattering, etc. and, in turn, on the reflection spectrum of the disk~\cite{Bambi:2013mha}.

In conclusion, the calculations of the thermal and reflection components require some assumptions about the astrophysical system (at least a model for the accretion disk, but the calculation of the reflection spectrum would also require a model for the corona) and some assumptions about fundamental physics (motion of photons and of particles around the compact object and atomic physics in strong gravitational fields). If the astrophysical part is well understood, we can think of testing the assumptions related to fundamental physics. Note that the Comptonized spectrum from the corona cannot be used to test fundamental physics, at least for the moment. The main reason is that we do not know the actual properties of the corona (geometry, temperature, location). The analysis of the thermal and reflection spectra can instead test fundamental physics because the properties of geometrically thin and optically thick accretion disks are thought to be well understood.

In the last few years, my group at Fudan University has developed two models for testing fundamental physics with black hole X-ray data: {\tt relxill\_nk}~\cite{Bambi:2016sac,Abdikamalov:2019yrr} and {\tt nkbb}~\cite{Zhou:2019fcg}. {\tt relxill\_nk} is an extension of the {\tt relxill} package developed by Thomas Dauser and Javier Garcia~\cite{Dauser:2013xv,Garcia:2013oma,Garcia:2013lxa} and can calculate reflection spectra of accretion disks in stationary, axisymmetric, and asymptotically flat spacetimes. {\tt nkbb} is instead a model for thermal spectra of accretion disks  in stationary, axisymmetric, and asymptotically flat spacetimes. Both models are public and available on GitHub\footnote{\url{https://github.com/ABHModels}}. They can be used with standard X-ray data analysis packages like {\tt XSPEC}.

%%%%%%%%%%%%%%%%%%%%%%%%%%%%%%%%%%%%%%%%%%

\section{Observational Constraints}\label{sec-oc}

There are two main strategies to test fundamental physics (in our case with black hole X-ray data, but actually this is true in general). These two strategies are normally referred to as {\it top-down} (or theory-specific) approach and {\it bottom-up} (or theory-agnostic) approach.

The top-down strategy is the most natural and logical one. In this case, we want to test some specific theory of gravity against general relativity. We can thus construct an astrophysical model for general relativity and another astrophysical model for the other theory of gravity, we analyze some astrophysical data with the two models, and eventually we use some statistical tool to check if one of the two models can explain the data better than the other model (or, in the most common case in which we want to test an extension of general relativity that includes general relativity in some special limit, we can measure and constrain the values of the parameters of the theory). The main drawback of this method is that it requires to know well the predictions of the theory to test, but this is not so easy. For example, we may want to test a theory of gravity in which the spacetime metric of astrophysical black holes is expected to be different from the Kerr solution of general relativity. In such a case, we need to know the rotating black hole solution of that theory, but in most cases we only know the non-rotating black hole solutions of theories beyond general relativity simply because it is easier to find a static and spherically symmetric solution rather than a stationary and axisymmetric one\footnote{Note that this is true even in general relativity. The Schwarzschild solution describing static and spherically symmetric black holes in general relativity was found by Schwarzschild a few months after Einstein had presented his theory, while the Kerr solution was found by Kerr only in 1963~\cite{Kerr:1963ud}.}.

In the bottom-up (or agnostic) approach, we parametrize possible deviations from the predictions of general relativity. For example, if we want to test the prediction that the spacetime metric around astrophysical black holes is described by the Kerr solution, we can analyze astrophysical data with a model employing a black hole metric in which some {\it deformation parameters} quantify possible deviations from the Kerr geometry and we can recover the Kerr solution for specific values of these deformation parameters. From the analysis of astrophysical data we can infer the values of these deformation parameters and check whether they are consistent with the Kerr solution. If, on the other hand, we want to test the value of the fine structure constant in the strong gravitational field of black holes, we can analyze astrophysical data with a model employing synthetic reflection spectra calculated for different values of $\alpha$. From the analysis of observations with this model, we can estimate the value of the fine structure constant near black holes. The main drawback of the bottom-up strategy is that normally we do not have a sufficiently general model to take into account any possible deviations from the predictions of general relativity. For example, we can use a certain deformed Kerr metric to test the Kerr hypothesis, but we do not have the most general black hole metric that can describe any possible deviation from the Kerr solution.

Up to now, most X-ray tests of general relativity with black holes have tested the Kerr hypothesis, namely if the spacetime metric around astrophysical black holes is described by the Kerr solution. In the literature, we can find examples of theory-agnostic tests as well as of theory-specific tests. Tests beyond the Kerr hypothesis are somewhat more complicated and may be explored in the next years. For example, in Ref.~\cite{Roy:2021pns} we tested the Weak Equivalence Principle with the bottom-up approach considering the possibility that either the motion of X-ray photons or the motion of the particles in the disk can deviate from the geodesic motion in the Kerr metric and we constrained possible violations of the Weak Equivalence Principle with a \textsl{NuSTAR} observation of the stellar-mass black hole in EXO~1846--031. A very preliminary study to use the analysis of the reflection spectrum to measure the value of the fine structure constant in the strong gravitational field of black holes was reported in Ref.~\cite{Bambi:2013mha} but without deriving any constraint from observations.

\subsection{Agnostic Tests of the Kerr Hypothesis}

Since 1960s, many experiments in the Solar System have tested the Schwarzschild solution in the weak field regime with the agnostic method. One can write the most general static and spherically symmetric line element as an expansion in $M/r$, where $M$ is the mass of the central object and $r$ is some radial coordinate. In isotropic coordinates, such a line element should read
\begin{eqnarray}\label{eq-ppn}
ds^2 = - \left( 1 - \frac{2M}{r} + \beta \frac{2M^2}{r^2} + ... \right) dt^2 + 
\left( 1 + \gamma \frac{2M}{r} + ... \right) \left( dx^2 + dy^2 + dz^2 \right)
\end{eqnarray}
in order to recover the correct Newtonian limit. In Eq.~(\ref{eq-ppn}), $\beta$ and $\gamma$ are unknown parameters to be determined by observations. If we write the Schwarzschild metric in isotropic coordinates, we find that $\beta=\gamma=1$. We can analyze experiments in the Solar System employing the metric in Eq.~(\ref{eq-ppn}) to measure $\beta$ and $\gamma$. As of now, Solar System experiments can confirm that the value of these two parameters is 1 with a precision at the level of $10^{-5}$, which confirms the Schwarzschild solution in the weak field regime within the precision of current experiments~\cite{Will:2014kxa}.

With a similar spirit, we can try to test the Kerr solution around astrophysical black holes. Unfortunately, we cannot use an expansion in $M/r$ because we want to probe the strong gravity region where $M/r$ is not a small parameter. At least for the moment, we do not have a general framework as in the case of Solar System experiments. In the literature, there are a number of {\it parametric black hole spacetimes} specifically proposed to test the Kerr hypothesis with electromagnetic data; see, for instance, Refs.~\cite{Johannsen:2011dh,Johannsen:2013szh,Cardoso:2014rha,Lin:2015oan,Konoplya:2016jvv,Ghasemi-Nodehi:2016wao,Mazza:2021rgq} for a non-complete list of options. Every proposal has its advantages and disadvantages. One of these parametric black hole spacetimes is the Johannsen metric~\cite{Johannsen:2013szh}, which has been extensively used for testing general relativity with X-ray data as well as with other techniques. In Boyer-Lindquist-like coordinates, the line element of the Johannsen metric is 
\begin{eqnarray}\label{eq-jp}
ds^2 &=&-\frac{{\Sigma}\left(\Delta-a^2A_2^2\sin^2\theta\right)}{B^2}dt^2
+\frac{{\Sigma}}{\Delta A_5}dr^2+{\Sigma} d\theta^2 \nonumber\\
&& +\frac{\left[\left(r^2+a^2\right)^2A_1^2-a^2\Delta\sin^2\theta\right]{\Sigma}\sin^2\theta}{B^2}d\phi^2 \nonumber\\
&& -\frac{2a\left[\left(r^2+a^2\right)A_1A_2-\Delta\right]{\Sigma}\sin^2\theta}{B^2}dtd\phi \, , 
\end{eqnarray}
where
\begin{eqnarray}
\Sigma &=& r^2 + a^2 \cos^2\theta \, , \\
\Delta &=& r^2 - 2 M r + a^2 \, , \\
B &=& \left(r^2+a^2\right)A_1-a^2A_2\sin^2\theta
\end{eqnarray}
and the functions $f$, $A_1$, $A_2$, and $A_5$ are defined as
\begin{eqnarray}
f &=& \sum^\infty_{n=3} \epsilon_n \frac{M^n}{r^{n-2}} \, , \qquad
A_1 = 1 + \sum^\infty_{n=3} \alpha_{1n} \left(\frac{M}{r}\right)^n \, , \\
A_2 &=& 1 + \sum^\infty_{n=2} \alpha_{2n} \left(\frac{M}{r}\right)^n \, , \qquad
A_5 = 1 + \sum^\infty_{n=2} \alpha_{5n} \left(\frac{M}{r}\right)^n \, .
\end{eqnarray}
The Johannsen metric has four infinite sets of deformation parameters: $\{ \epsilon_n \}$ ($n=3,4,...$), $\{ \alpha_{1n} \}$ ($n=3,4,...$), $\{ \alpha_{2n} \}$ ($n=2,3,...$), and $\{ \alpha_{5n} \}$ ($n=2,3,...$). This form of the Johannsen metric has the correct Newtonian limit and cannot be distinguished from the Schwarzschild metric by Solar System experiments (for this reason there are not $\epsilon_1$, $\epsilon_2$, $\alpha_{11}$, $\alpha_{12}$, $\alpha_{21}$, and $\alpha_{51}$). $\alpha_{13}$ is the leading order correction to the Kerr spacetime as well as the parameter with the strongest impact on the spectra of thin disks, and for this reason most tests of the Kerr hypothesis have focused on the constraints on the value of this parameter.

\begin{figure}[t]
\centering
\includegraphics[width=0.85\textwidth,trim=0.2cm 0.2cm 0cm 0cm,clip]{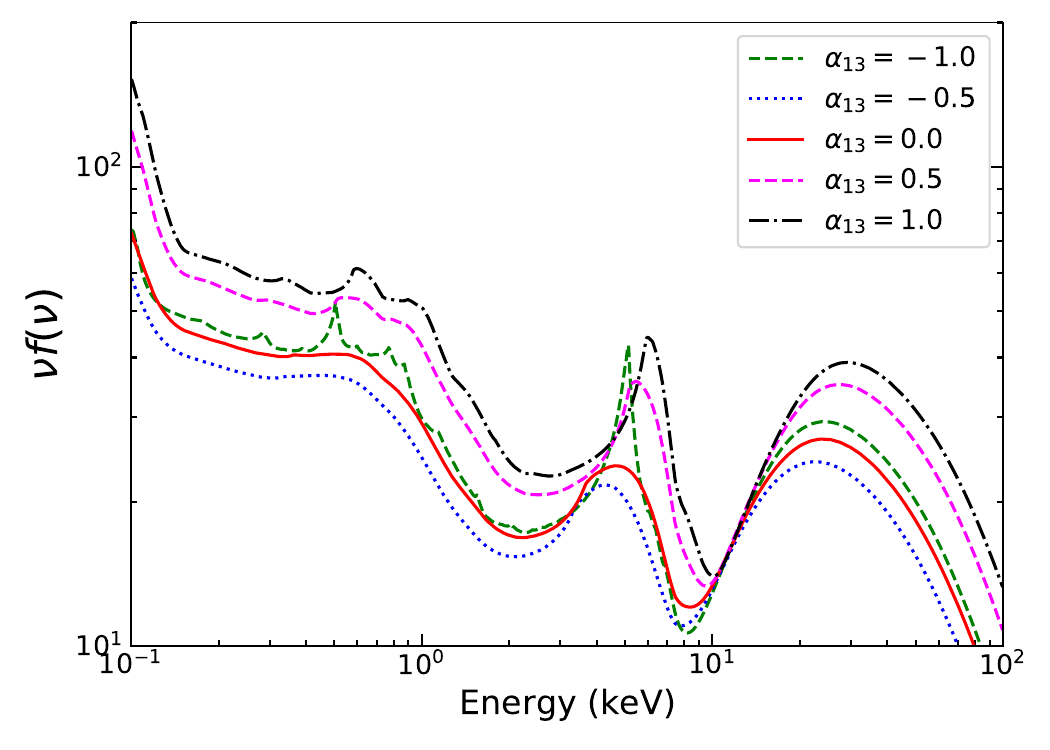}
\caption{Synthetic reflection spectra of thin disks in the Johannsen spacetime for different values of the deformation parameter $\alpha_{13}$. The values of the other model parameters are not changed. Figure from Ref.~\cite{Bambi:2021chr} under the terms of the Creative Commons Attribution 4.0 International License.}
\label{f-relxillnk}
\end{figure}

\begin{figure}[t]
\centering
\includegraphics[width=0.85\textwidth,trim=0cm 0cm 0cm 0cm,clip]{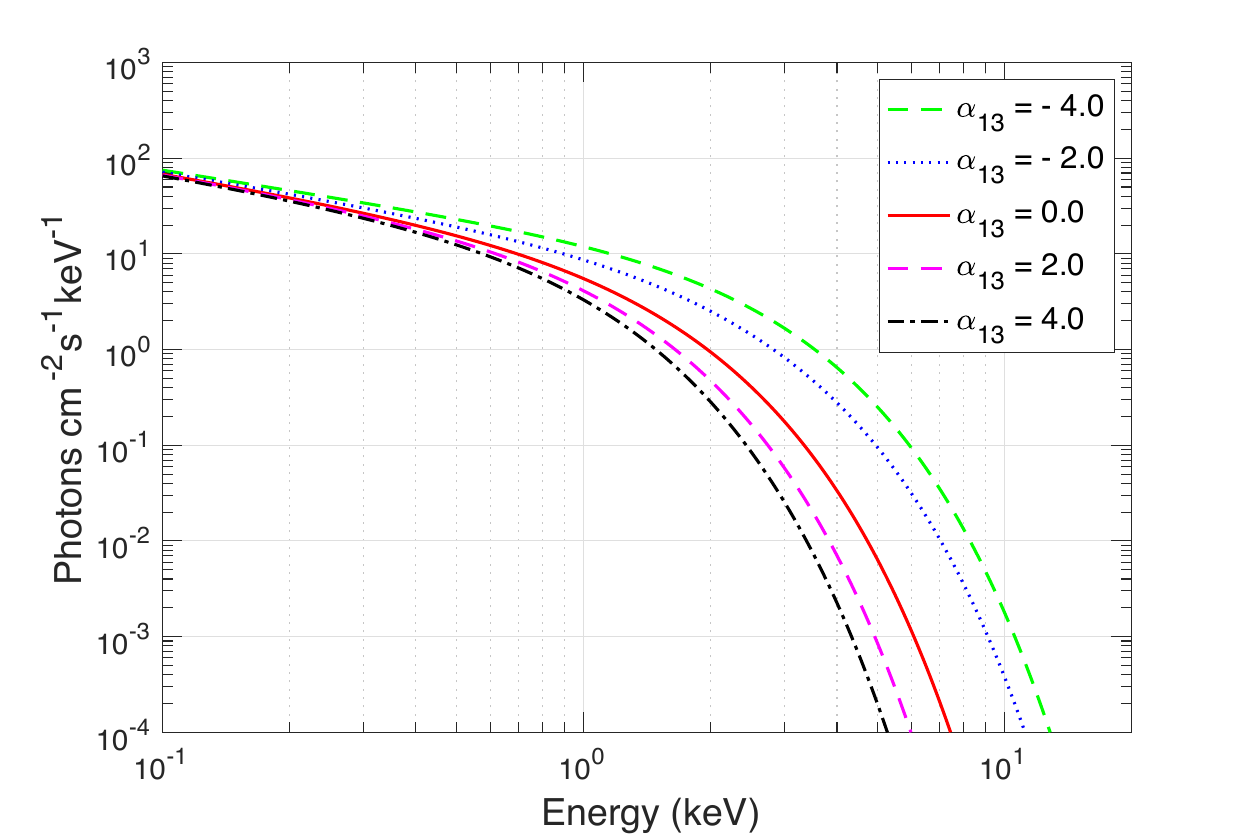}
\vspace{0.2cm}
\caption{Synthetic thermal spectra of thin disks in the Johannsen spacetime for different values of the deformation parameter $\alpha_{13}$. The values of the other model parameters are not changed. Figure from Ref.~\cite{Zhou:2019fcg}.}
\label{f-nkbb}
\end{figure}

Fig.~\ref{f-relxillnk} shows some reflection spectra of thin disks in the Johannsen spacetime for different values of the deformation parameter $\alpha_{13}$ calculated with {\tt relxill\_nk}. Fig.~\ref{f-nkbb} shows some thermal spectra of thin disks in the Johannsen spacetime calculated with {\tt nkbb}. In both cases, all the other parameters of the models are kept constants and different spectra are simply obtained by changing the value of the deformation parameter $\alpha_{13}$ (while all other deformation parameters of the Johannsen metric vanish). The red spectra with $\alpha_{13} = 0$ are those in the Kerr metric. As we can see from Fig.~\ref{f-relxillnk} and Fig.~\ref{f-nkbb}, the value of the deformation parameter $\alpha_{13}$ has a clear impact on the shape of the reflection and thermal spectra. It is thus clear that we can use {\tt relxill\_nk} and {\tt nkbb} to analyze X-ray spectra of accreting black holes and, modulo degeneracy with other parameters, we can measure $\alpha_{13}$.

\begin{figure}[t]
\centering
\includegraphics[width=0.90\textwidth,trim=3.5cm 0.0cm 3.5cm 1.0cm,clip]{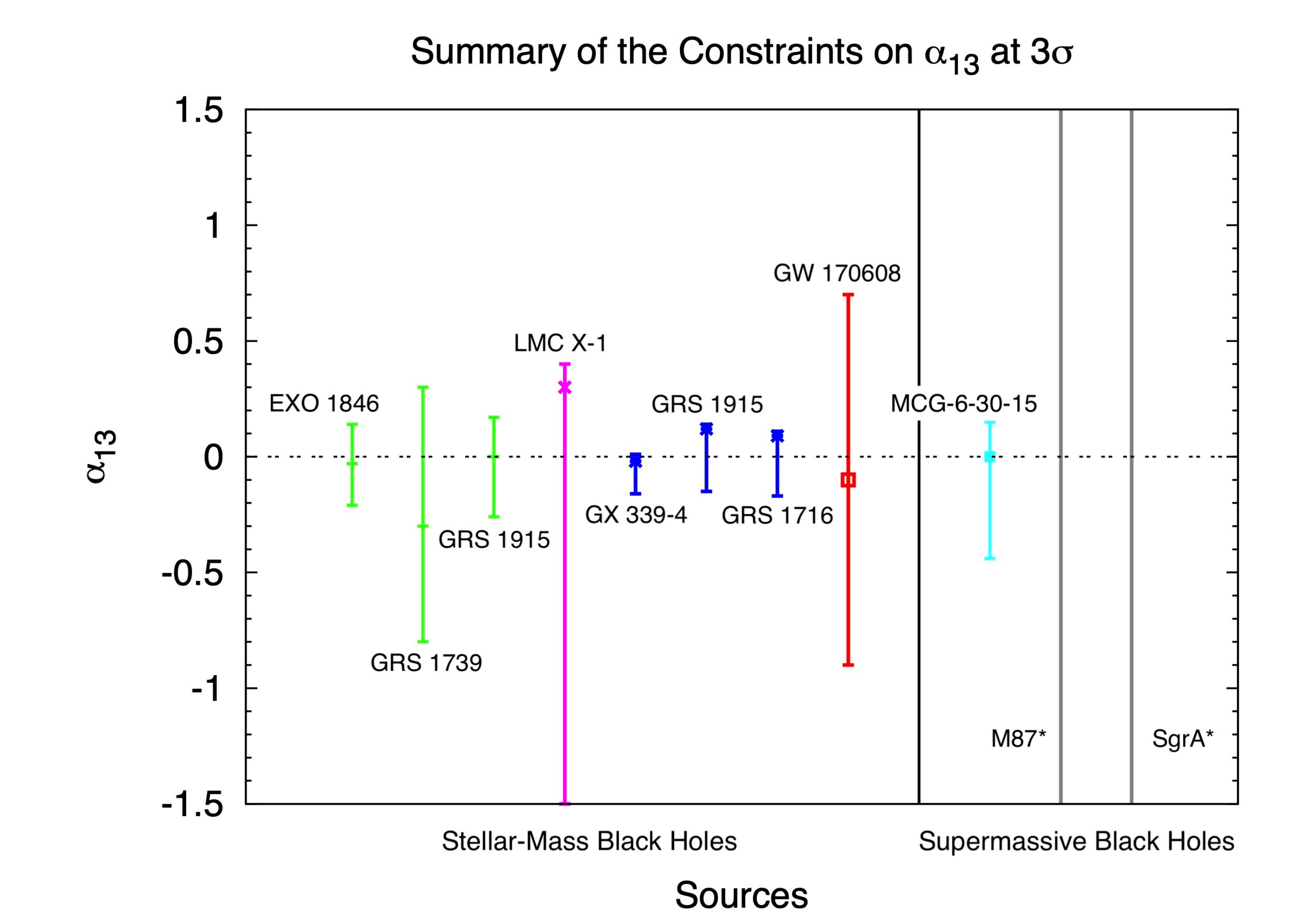}
\caption{Summary of current measurements of the Johannsen deformation parameter $\alpha_{13}$ with different techniques. Every measurement shows the best-fit value, the 3-$\sigma$ upper and lower uncertainties, and the name of the source. The constraints in green are the three most precise and accurate measurements of $\alpha_{13}$ for stellar-mass black holes from the analysis of the reflection features using {\tt relxill\_nk} (EXO~1846--031~\cite{Tripathi:2020yts}, GRS~1739--278~\cite{Tripathi:2020yts}, and GRS~1915+105~\cite{Zhang:2019ldz,Abdikamalov:2020oci}). The constraint in magenta is the measurement of $\alpha_{13}$ for the stellar-mass black hole in LMC~X-1 from the analysis of thermal spectra using {\tt nkbb}~\cite{Tripathi:2020qco}. The constraints in blue are obtained combining the analysis of the reflection features with {\tt relxill\_nk} and of the thermal spectrum with {\tt nkbb} for the same source (GX~339--4~\cite{Tripathi:2020dni}, GRS~1915+105~\cite{Tripathi:2021rqs}, and GRS~1716--249~\cite{Zhang:2021ymo}). The constraint in red is the most precise measurement of $\alpha_{13}$ for stellar-mass black holes with gravitational wave data~\cite{Shashank:2021giy}. The constraint in cyan is the most precise and accurate measurement of $\alpha_{13}$ for supermassive black holes from the analysis of reflection features using {\tt relxill\_nk} (MCG--6--30--15~\cite{Tripathi:2018lhx}). Last, the two constraints in gray are from black hole imaging for the supermassive black holes M87$^*$~\cite{EventHorizonTelescope:2020qrl} and Sgr~A$^*$~\cite{EventHorizonTelescope:2022xqj} (in both cases, the authors do not report the best-fit values but only the allowed $\alpha_{13}$ ranges, which are larger than the range $(-1.5,1.5)$ shown in this figure). The horizontal dotted line at $\alpha_{13}=0$ marks the Kerr solution of general relativity. Figure adapted from Ref.~\cite{cb22}.}
\label{f-summary}
\end{figure}

Fig.~\ref{f-summary} compares the current measurements of the Johannsen deformation parameter $\alpha_{13}$ obtained from the analysis of X-ray data with {\tt relxill\_nk} and {\tt nkbb}, from the analysis of gravitational wave data, and from black hole imaging. We distinguish tests of the Kerr hypothesis with stellar-mass black holes from those with supermassive black holes, as they can potentially probe different regimes. In the case of the constraints from reflection features and gravitational wave data, only the most precise and accurate measurements on $\alpha_{13}$ are shown in Fig.~\ref{f-summary}.

The most stringent constraints to date are from the simultaneous analysis of reflection features and thermal spectra of the stellar mass black holes in GX~339--4, GRS~1915+105, and GRS~1716--249 (blue data in Fig.~\ref{f-summary}). In general, the sole analysis of the thermal spectrum cannot constrain a deformation parameter well because the thermal spectrum has a very simple shape (see Fig.~\ref{f-nkbb}): even if we know the mass and the distance of the source, we cannot constrain simultaneously the spin, the mass accretion rate, and the deformation parameter (see the magenta constraint in Fig.~\ref{f-summary} and the discussion in Ref.~\cite{Tripathi:2020qco}). 
For supermassive black holes, the thermal spectrum is peaked in the UV band, where dust absorption prevents any accurate measurement and therefore we cannot use the analysis of the thermal spectrum and {\tt nkbb} to test supermassive black holes.
In general, we can obtain stronger constraints from stellar-mass black holes than from supermassive black holes (the sources are brighter and normally their spectra are not affected by absorption of material crossing our line of sight, even if their hotter disks are more difficult to model). However, MCG--6--30--15 is an exceptional case (cyan bar in Fig.~\ref{f-summary}): the source is very bright, its spectrum has often a very strong and broadened iron line, and the constraint on $\alpha_{13}$ is obtained from the analysis of high quality simultaneous observations \textsl{NuSTAR} and \textsl{XMM-Newton} that combine a broad energy spectrum and a high energy resolution in the iron line region. In the end, the constraint on $\alpha_{13}$ from MCG--6--30--15 is comparable to the best constraints from stellar-mass black holes with only {\tt relxill\_nk} (green data). The red constraint from gravitational wave data of the coalescence of two stellar-mass black holes is obtained following the approach proposed in Ref.~\cite{Cardenas-Avendano:2019zxd}, where one assumes that the gravitational wave emission is the same as in general relativity and the constraint on $\alpha_{13}$ only comes from the motion of the bodies; within an agnostic approach, it could not be otherwise, because here we only know the spacetime metric -- in this specific case, the Johannsen metric with the deformation parameter $\alpha_{13}$ -- and we do not know the field equations of the theory. These parametric black hole spacetimes are obtained by deforming the Kerr solution following certain criteria; they are not the exact solutions of specific gravity theories. The best gravitational wave constraint on $\alpha_{13}$ is weaker than the best X-ray constraints, but for other deformations from the Kerr solution we may find opposite results~\cite{Abdikamalov:2021zwv,Yu:2021xen,Shashank:2021giy}: this is understandable because X-ray and gravitational wave tests are sensitive to very different physics and different relativistic effects, so X-ray tests can constrain better some deformations from the Kerr geometry and gravitational wave tests can constrain better other deformations. Last, we have the constraints from M87$^*$ and Sgr~A$^*$ (gray data in Fig.~\ref{f-summary}) from black hole imaging. These constraints are definitively weaker than those from X-ray and gravitational wave tests. They may be able to reach the current sensitivity of X-ray and gravitational wave tests with the launch of a telescope to space: such a possibility is already discussed within the community and could improve the current angular resolution by an order of magnitude.

\subsection{Theory-Specific Tests of the Kerr Hypothesis}

{\tt relxill\_nk} has been used even to test specific theories of gravity in which electrically neutral black holes are not described by the Kerr solution. In the end, this technique is quite general. We just need to implement the correct black hole metric in {\tt relxill\_nk} and then analyze some reflection-dominated black hole spectra to measure the parameters of the theory.

In Einstein-Maxwell-dilaton-axion gravity, black holes are expected to have a dilaton charge $r_2 \ge 0$ and the Kerr solution of general relativity is recovered when $r_2 = 0$. In Ref.~\cite{Tripathi:2021rwb}, we analyzed a \textsl{NuSTAR} observation of the stellar-mass black hole in EXO~1846--031 and we inferred the following constraint on $r_2$
\begin{eqnarray}
r_2 < 0.011 \quad (\text{90\% CL}) \, .
\end{eqnarray}

Conformal gravity is a family of theories beyond general relativity that have been proposed to solve the problem of spacetime singularities~\cite{Bambi:2016wdn,Bambi:2016yne,Chakrabarty:2017ysw}. The singularity-free black holes in conformal gravity are characterized by a new parameter, $L > 0$. For $L=0$, we recover the singular Kerr solution of general relativity. From the analysis of black hole X-ray data with {\tt relxill\_nk} we can infer the following constraint on $L/M$ (where $M$ is the black hole mass)~\cite{Zhou:2019hqk,Zhou:2018bxk}
\begin{eqnarray}
L/M < 0.09 \quad (\text{90\% CL}) \, .
\end{eqnarray}

If we quantize general relativity, we obtain an effective quantum field theory valid only at energies much lower than the Planck scale. Asymptotically safe quantum gravity is a promising candidate scenario to provide a UV extension for such an effective field theory of general relativity. In Ref.~\cite{Held:2019xde}, the authors proposed a rotating black hole metric in asymptotically safe quantum gravity. Such a solution is characterized by a new parameter, $\tilde{\gamma} > 0$, which is inversely proportional to the asymptotically safe fixed-point value of the theory. The Kerr metric is recovered when $\tilde{\gamma} = 0$. In Ref.~\cite{Zhou:2020eth}, we implemented such a rotating black hole metric in asymptotically safe quantum gravity in {\tt relxill\_nk} and we analyzed a \textsl{Suzaku} observation of the stellar-mass black hole in GRS~1915+105. From our analysis, we measured $\tilde{\gamma}$
\begin{eqnarray}
\tilde{\gamma} < 0.047 \quad (\text{90\% CL}) \, .
\end{eqnarray}
The constraint on $\tilde{\gamma}$ from X-rays is much stronger than that obtained from black hole imaging in Ref.~\cite{Held:2019xde}.

Even if we assume that general relativity is the correct theory of gravity, there is not yet a proof that the final product of the complete gravitational collapse of an uncharged body is a Kerr black hole. On the other hand, today we know exact solutions of the Einstein Equations in which the complete collapse of a body does not produce a Kerr black hole and instead leads to a spacetime with naked singularities~\cite{Joshi:2011rlc}. In Ref.~\cite{Tao:2023hou} we explored the possibility that the spacetime around gravitationally collapsed objects is described by the $\delta$-Kerr metric~\cite{deltakerr1,deltakerr2}, which is an exact solution of the Einstein Equations that can be obtained from a non-linear superposition of the $\delta$-metric and the Kerr metric. The $\delta$-Kerr metric has an extra parameter, $q$, which measures deviations from the Kerr solution. The Kerr metric is recovered for $q=0$, while the object is more oblate (prolate) than a Kerr black hole if $q > 0$ ($q < 0$). Implementing the $\delta$-Kerr metric in {\tt relxill\_nk} and analyzing the \textsl{NuSTAR} spectrum of the stellar-mass black hole in EXO~1846--031, in Ref.~\cite{Tao:2023hou} we inferred the following constraint on $q$
\begin{eqnarray}
-0.1 < q < 0.7 \quad (\text{90\% CL}) \, .
\end{eqnarray}

The examples reported in this sub-section are to show that {\tt relxill\_nk} is a tool that can normally test any theory of gravity in which we know the rotating black hole solution and it is different from the Kerr metric. We also note that the constraints inferred from the analysis of reflection spectra with {\tt relxill\_nk} are more stringent than the constraints inferred with other electromagnetic techniques. On the other hand, constraints from gravitational wave data are difficult to infer because they require precise calculations of gravitational waveforms in those theories, which are normally not well studied yet.

%%%%%%%%%%%%%%%%%%%%%%%%%%%%%%%%%%%%%%%%%%

\section{Accuracy of X-ray Tests of General Relativity}

Are the X-ray tests of the Kerr hypothesis presented in the previous section {\it robust}? From Fig.~\ref{f-summary}, we see that the X-ray measurements of the Johannsen deformation parameter $\alpha_{13}$ are certainly more {\it precise} than those with current gravitational wave data and black hole imaging, but are they also {\it accurate}?

The answer to this question is that we believe that these tests of the Kerr hypothesis with black hole X-ray data are robust, but we must be very careful to select the right sources and the right observations. The point is that we do not need to test as many sources as possible to get a large number of (accurate and inaccurate) measurements. We can instead focus on a few number of spectra that are thought to be well understood and can provide very precise measurements. Let us note that the situation is very different from most astrophysical studies, where, for example, we want to measure the spin parameter of as many sources as possible to study the spin distribution of a whole black hole population. Another example is the case in which we want to study a certain kind of accretion state of black holes, even if such a state is not ideal for precise measurements of the system. The situation is also very different from the tests with gravitational wave data and black hole imaging. In the case of gravitational wave data, all systems are quite similar and share the same systematic uncertainties in the measurement of their parameters (in the end, they are just two black holes in vacuum!): this is also evident from the fact that the constraints from different sources on possible deviations from the Ker solution are all quite similar~\cite{Cardenas-Avendano:2019zxd,Shashank:2021giy}, while in the case of X-ray tests we can have sources that provide very strong constraints and sources that cannot provide constraints at all. In the case of black hole imaging, we have only M87$^*$ and Sgr~A$^*$, as the angular size of any other supermassive black hole in our sky is too small, and therefore we have to understand those sources and their observations well even if they are not ideal for tests of fundamental physics.

Generally speaking, our tests with X-ray data require to select sources in which $a)$ the signature of relativistic effects is strong in the spectrum, and $b)$ the accretion process is well understood and constrained. Point $a)$ is necessary to break the degeneracy between possible new physics and the astrophysical model, as well as to have precise measurements/constraints of new physics. Point $b)$ is required to limit the systematic uncertainties of the astrophysical model. To be more specific, the sources and observations suitable for testing fundamental physics with the analysis of the reflection features should meet the following requirements:
\begin{enumerate}  
\item The inner edge of the accretion disk should be as close as possible to the compact object. A necessary but not sufficient condition is that the spin parameter of the compact object is high (say $a_* > 0.9$ if it is a Kerr black hole). 
\item The corona should be compact and close to the compact object.
\item The accretion disk should be geometrically thin and optically thick, with the inner edge at or near the innermost stable circular orbit (or ISCO). A necessary but not sufficient condition is that the Eddington-scaled accretion luminosity is between a few percent and about 30\%. 
\item The spectrum must have a prominent iron line.
\item The source must be bright (and the data should not be affected by pile-up).
\item The geometry of the accretion disk and of the corona should not change during the observation.
\item The X-ray data should cover both the iron line region and the Compton hump, and possibly have a good energy resolution in the iron line region.
\end{enumerate}

Points~1 and 2 are necessary to have most of the reflection component generated very close to the black hole, so that the relativistic effects in the X-ray spectrum are stronger and thus they are more easy to measure with good precision; see also Refs.~\cite{Bambi:2022dtw,Dauser:2013xv,Kammoun:2018ltv}. Since {\tt relxill\_nk} assumes that the accretion disk is thin and the motion of the gas is Keplerian, we must select the sources with such a disk, which is point~3 above. Simulations of reflection spectra from GRMHD-generated thin disks show that we can recover the correct input parameters well from the X-ray spectrum~\cite{Shashank:2022xyh}. On the contrary, if we analyze sources with a too high mass accretion rate, the disk is thick, and we can easily get very precise, but inaccurate, measurements of the spacetime metric~\cite{Riaz:2019bkv,Riaz:2019kat}. Since the shape of the iron K$\alpha$ line is the most informative part of the reflection spectrum for our tests, it is important to analyze data with prominent and very broadened iron lines, which is point~4 above. Note, for example, that in a fully ionized disk the iron line disappears in the X-ray spectrum of the source. If we select the right sources and the right observations, the uncertainty in the final measurement is dominated by the statistic uncertainty in the photon count, which decreases if the source is bright (point~5). If the geometry of the accretion flow in the strong gravity region or the geometry of the corona change during the observation, this should be taken into account in the data analysis process and it is definitively a complication. If the properties of the disk and the corona do not change, point~6, this simplifies the analysis. 
Last, it is certainly useful both to have a good energy resolution in the iron line region (because the shape of the iron line is the most informative part of the spectrum) and to analyze a broad spectrum including the Compton hump (because this helps to constrain other parameters of the model, like the ionization of the disk, and, in turn, even to get a better measurement of the spacetime metric around the black hole), which is point~7. For this reason, the constraints reported in Section~\ref{sec-oc} are mainly obtained from \textsl{NuSTAR} data (which permit to fit the energy band 3-80~keV) or, like in the case of MCG--6--30--15, simultaneous observations \textsl{NuSTAR} and \textsl{XMM-Newton} (which permit also to have a good energy resolution in the iron line region).

%%%%%%%%%%%%%%%%%%%%%%%%%%%%%%%%%%%%%%%%%%

\section{Concluding Remarks}

This article has summarized the state-of-the-art of tests of general relativity with black hole X-ray data. In the past few years, we have developed the reflection model {\tt relxill\_nk} and the thermal model {\tt nkbb} to test fundamental physics from the analysis of X-ray spectra of accreting black holes. So far, we have mainly worked on tests of the Kerr hypothesis; that is, to test if the spacetime metric around astrophysical black holes is described by the Kerr solution as expected in the framework of standard physics (general relativity, no exotic matter fields, no naked singularities, etc.). In the future, these tests can be extended to study possible deviations from geodesic motion induced by new fields, violation of Lorentz Invariance, new interactions between the gravity and the matter sectors, etc. While X-ray tests of general relativity are not as popular as tests with gravitational wave data or black hole imaging, they can provide very competitive constraints, usually somewhat stronger than those possible with current gravitational wave observations and certainly much stronger than those with current images of the supermassive black holes M87$^*$ and Sgr~A$^*$.

If we select carefully the sources and the observations, X-ray tests of the Kerr hypothesis can be robust, and in such a case the constraints on possible deviations from the Kerr solution are limited by the quality of the data. To improve significantly current X-ray constraints we have to wait for the next generation of X-ray missions, starting from \textsl{eXTP}~\cite{eXTP:2016rzs} (currently scheduled to be launched in 2028). For example, an observation with \textsl{eXTP} can roughly improve the constraint on a deformation parameter by an order of magnitude with respect to a similar observation with \textsl{NuSTAR} of the same source~\cite{Tao:2023hou}. However, in order to fully exploit the high-quality data from the next generation of X-ray missions, it will be necessary to develop more sophisticated theoretical models than those available today or, otherwise, we risk to have very precise, but not very accurate, measurements of the spacetime metric.  
Future models should include the effect of the returning radiation (i.e. the radiation emitted by the disk and returning to the disk because of the strong light bending near black holes), more accurate calculations of the reflection spectrum in the rest-frame of the material in the disk, more sophisticated descriptions of the emissivity profiles by implementing specific coronal geometries, and more sophisticated accretion disk models than the Novikov-Thorne one (probably employing GRMHD-generated accretion disks).

Gravitational wave tests promise to improve quickly in the near future and we can expect that eventually they will be able to put the most stringent constraints on the Kerr hypothesis (see, for instance, Ref.~\cite{Babak:2017tow}). However, X-ray tests can still be interesting because they are complementary to the gravitational wave ones. X-ray tests can indeed probe better the interactions between the gravity and the matter sector. For example, non-minimal coupling between the gravity and the matter sectors could induce deviations from geodesic motions of photons or some particle species, variation of fundamental constants in the strong gravity region around black holes, etc. These phenomena are more difficult or impossible to detect with gravitational wave data. On the other hand, gravitational wave tests are certainly more suitable to probe the gravitational sector itself and, in particular, are sensitive to the dynamical regime, which cannot be probed by electromagnetic tests.

%%%%%%%%%%%%%%%%%%%%%%%%%%%%%%%%%%%%%%%%%%
\vspace{6pt} 

\funding{This work was supported by the National Natural Science Foundation of China (NSFC), Grant No.~11973019, 12250610185, and 12261131497, the Natural Science Foundation of Shanghai, Grant No.~22ZR1403400, the Shanghai Municipal Education Commission, Grant No.~2019-01-07-00-07-E00035, and Fudan University, Grant No.~JIH1512604.}

\dataavailability{The models {\tt relxill\_nk} and {\tt nkbb} are available on GitHub at \url{https://github.com/ABHModels}. Tests of the Kerr hypothesis with {\tt relxill\_nk} and {\tt nkbb} have used data available on the HEASARC Data Archive website at \url{https://heasarc.gsfc.nasa.gov/docs/archive.html}.} 

%\acknowledgments{In this section you can acknowledge any support given which is not covered by the author contribution or funding sections. This may include administrative and technical support, or donations in kind (e.g., materials used for experiments).}

%%%%%%%%%%%%%%%%%%%%%%%%%%%%%%%%%%%%%%%%%%

\begin{adjustwidth}{-\extralength}{0cm}

\reftitle{References}

% If authors have biography, please use the format below
%\section*{Short Biography of Authors}
%\bio
%{\raisebox{-0.35cm}{\includegraphics[width=3.5cm,height=5.3cm,clip,keepaspectratio]{Definitions/author1.pdf}}}
%{\textbf{Firstname Lastname} Biography of first author}
%
%\bio
%{\raisebox{-0.35cm}{\includegraphics[width=3.5cm,height=5.3cm,clip,keepaspectratio]{Definitions/author2.jpg}}}
%{\textbf{Firstname Lastname} Biography of second author}

% For the MDPI journals use author-date citation, please follow the formatting guidelines on http://www.mdpi.com/authors/references
% To cite two works by the same author: \citeauthor{ref-journal-1a} (\citeyear{ref-journal-1a}, \citeyear{ref-journal-1b}). This produces: Whittaker (1967, 1975)
% To cite two works by the same author with specific pages: \citeauthor{ref-journal-3a} (\citeyear{ref-journal-3a}, p. 328; \citeyear{ref-journal-3b}, p.475). This produces: Wong (1999, p. 328; 2000, p. 475)

%%%%%%%%%%%%%%%%%%%%%%%%%%%%%%%%%%%%%%%%%%
%% for journal Sci
%\reviewreports{\\
%Reviewer 1 comments and authors’ response\\
%Reviewer 2 comments and authors’ response\\
%Reviewer 3 comments and authors’ response
%}
%%%%%%%%%%%%%%%%%%%%%%%%%%%%%%%%%%%%%%%%%%
\PublishersNote{}
\end{adjustwidth}
\end{document}